# Raman spectra of electrochemically hydrogenated diamond like carbon surface

Hari Shankar Biswas[a,b], Jagannath Datta[a,c], Pintu Sen[d], Uday Chand Ghosh[e], Nihar Ranjan Ray[a,*]

[a]Nanocrystalline Diamond like Carbon Synthesis Laboratory, Saha Institute of Nuclear Physics 1/AF, Bidhan Nagar, Kolkata-700 064 India
[b]Assistant Professor in Chemistry, Department of Chemistry, Surendranath College,24/2 Mahatma Gandhi road, Kolkata-700009, India
[c]Analytical Chemistry Division,Bhabha Atomic Research Centre, Variable Energy Cyclotron Centre, 1/AF, Bidhan Nagar, Kolkata-700 064, India
[d]Physics Department, Variable Energy Cyclotron Center, 1/AF, Bidhannagar, Kolkata-700 064, India
[e]Department of Chemistry, Presidency University, 86/1 College Street, Kolkata-700 073 , India

*Corresponding author: niharranjan.ray@saha.ac.in

Abstract

Raman spectroscopy has been employed to distinguish between the Raman spectrum of pristine hydrogenated diamond like carbon (PHDLC) and that of electrochemically hydrogenated diamond like carbon (ECHDLC). The enhancement of the background photoluminescence (PL) in the Raman spectrum and broadening of PL spectrum of ECHDLC are identified to be due to increase of $sp^3$ C-H density onto the PHDLC surface, during novel electrochemical process of hydrogenation of $sp^2$ C=C into $sp^3$ C-H.

1. Introduction

P-type surface conductivity (SC) of hydrogenated diamond (HD) films [1-3] is unique among semiconductors. Experiments performed in UHV and in air revealed that, in addition to the surface hydrogenation, exposure to air was a necessary condition for the SC; the SC disappears after dehydrogenation or oxidation of HD surface [4]. The experimental results [4] confirm that presence of C-H stretching modes on the HD surface is necessary for the high surface conductivity in (~$10^{-6}$ $\Omega^{-1}$) in HD. It is argued in the literatures [1-4] that the $H_2O$ or the $H_3O^+$ adlayer can act as an electrostatic surface dipole, creating hole accumulation on the surface and resulting in an electrostatic potential step, called barrier, developed on the HD surface. Depending upon distribution of holes (acceptors) accumulation, caused by hydrogenation, two-dimensionally on the surface or three-dimensionally from the surface into the subsurface region, the mechanism of SC of p-type unipolar HD surface will be influenced [5].

We have demonstrated earlier [6] that the coherency of diamond like $sp^3$ C–H and graphite like $sp^2$ C = C carbons can produce a continuous nonporous hydrogenated diamond like carbon (HDLC) thin film (thickness ~ 168 nm) having atomically smooth surface; this pristine HDLC (PHDLC) film, grown onto large area substrate Si (100), has both p-type and n-type surface conduction of non-linear type, but the annealed sample shows surface conductance of ohmic nature [7]. SC of PHDLC does not need water addlayer; also SC does not disappear due to dehydrogenation [7] as in the case HD [1-5]. Moreover, SC of PHDLC is ambipolar but SC of HD is p-type unipolar. Thus we have described, tuning of electronic properties of PHDLC surface should be possible by controlling the hydrogenation/dehydrogenation of PHDLC surface [7].

There are numerous techniques in the literatures for the hydrogenation of $sp^2$ C = C carbons, viz., exposure to atomic hydrogen source [8-10], electron beam exposure of a water adhesive layer on graphene [11], exposure of reactive ion etching plasma on graphene [12]. We know that depending upon the chemical potential of electrons in the liquid phase ( $\mu_e$ ) and in diamond ( Fermi level $E_F$ ) the redox reaction is driven towards hydrogenation of diamond surface in contact with a water layer [4,13]. The motivation in the present work is to explore the occurrence of the similar redox reaction during the electrochemical process on the PHDLC surface in contact with suitable electrolyte. In this paper we report, for the first time, the novel hydrogenation of PHDLC surface through electrochemical process, as evident from the typical Raman spectra.

2. Experimental

   The synthesis of PHDLC film is described elsewhere [14]. Electrochemical behavior of PHDLC film in 1M KCl (pH = 7) through cyclic voltametry (CV) measurement was investigated with AUTOLAB-30 potentiostat/galvanostat. A platinum electrode and a saturated Ag/AgCl electrode were used as counter and reference electrodes respectively. Cyclic voltammograms were recorded between -1v to +1v w.r.t. with PHDLC film as reference electrode. The typical scan rate was 20mV/s. A specially constructed electrochemical cell, wherein PHDLC and Pt discs are two electrodes with exposed surface of 1cm in diameter of each and separated by 5mm teflon sheet, in 1M KCl (pH=1). This cell was used for hydrogenation of PHDLC surface. A schematic diagram of the cell is shown in figure 1. Raman spectra were obtained using a confocal micro Raman spectrometer (Lab RAM HR Vis., Horiba Jobin Yvon S.A.S. France). Under the operating conditions of the spectrometer as described elsewhere [6], we get the Raman spectrum of bulk of the sample. A Keithley 2635 source meter was used to measure the sheet resistance of sample surface.

3. Results and Discussion

   A typical CV measurement of PHDLC film in 1M KCl (pH=7) is shown in figure 2a. The measured values are: peak cathodic potential $E_{pc}$ ~ +0.214v and peak anodic potential $E_{pa}$ ~ -0.208v; the peak separation ($\Delta E=0.0592/n$, n is the number of electrons transferred) is nearly 6 mV. The measured typical value of peak separation ( ~ 6mV) corresponds to a reversible cyclic voltammogram with n=10 at the scan rate 20mV/s; also the current peak ratio ($i_{pa}/i_{pc}$) very close to 1 clearly indicates reversibility of redox reaction [15]. Thus the PHDLC film is capable of undergoing nearly reversible surface redox couple reaction [15]. Hence our PHDLC film can donate (oxidation) as well as accept (reduction) electrons like noble metal Pt. This observation seems to be reasonable due to the presence of $sp^2$ C=C (donor) and $sp^3$ C-H (acceptor) carbons coherently in the film [ 6 ]. The typical open circuit potential difference ($PD_{open}$) between PHDLC and Pt electrodes in 1M KCl (pH=1) electrolyte of the electrochemical cell (see fig.1) is ~ -0.650v. This observation clearly indicates that $\mu_e$ is above $E_F$ and thus electrons are being transferred from liquid to PHDLC, which in turn becomes negatively charged in equilibrium ($E_F = \mu_e$). A schematic picture of the electron transfer process between PHDLC surface and electrolyte is shown in figure 2b. This picture is completely opposite to that in the case of HD [4]. Due to high negative electron affinity ($\chi_{C:H}$) and band gap ($E_g$ ~ 4.5eV) of PHDLC [7], the electron affinity of the adsorbate $\chi_{ad}$ ( = $E_g$ + $\chi_{C:H}$ ) is negative; thus electrons will be transferred from electrolyte to PHDLC. When we close the electrodes of the cell in fig.1, via an external resistance, say 1KΩ, for a few hours, it is observed that the equilibrium potential difference between the electrodes comes down to smaller value ~ - 0.015v. During close circuit condition, it is possible that some $H^+$ ions moving from electrolyte towards the negatively charged PHDLC surface, convert some $sp^2$ C=C into $sp^3$ C-H carbons and thus the value of $PD_{open}$ is observed to decrease gradually. We may now call the PHDLC as ECHDLC. The typical sheet resistances of PHDLC and ECHDLC surfaces are measured to be 75MΩ ( for sourcing current 100 nA) and 5-8 GΩ( for sourcing current 1000 nA) respectively. This result signifies permanent increment of hydrogen content via conversion of $sp^2$ C=C into $sp^3$ C-H carbons onto the surface of PHDLC, now called as ECHDLC. The typical Raman spectra and PL spectra of PHDLC and ECHDLC are shown in figs.3a-c. The main effect of hydrogen incorporation in the PHDLC surface is to modify its C-C network [16]. Instead of increasing the fraction of C-C bond, hydrogen converts C=C into C-H bond to increase the $sp^3$ C-H content in the film. $sp^3$ C-H content in the film can be estimated using the formula $sp^3$ content = 0.24-48.9 ( $\omega_G$ – 0.1580), where $\omega_G$ is the position of G-peak in unit of inverse micrometer unit [17]. As a result of the recombination of electron-hole pairs within $sp^2$ C=C and $sp^3$ C-H carbon matrix in the film, a linear photoluminescence (PL) background to the Raman spectrum appears [18]. The ratio between the slope m of the fitted linear background of the Raman spectrum (shown by dotted line in fig.3a & b) due to PL and the intensity of G peak ( $I_G$), m/ $I_G$, can be used as a measure of the bonded H content in the film [17,18]. The slope m is described in micrometer unit. Also broadening of PL spectrum is a strong signature of increase of bonded hydrogen in the film [18]. Using this analysis, the effect of increase of bonded hydrogen content on the electrochemically hydrogenated PHDLC surface is evident from the significant increase of PL background in the sample

ECHDLC, as shown by dotted line in fig.3a&b respectively; also the same effect is evident in the broadening of PL spectrum, as shown in fig.3c. The $sp^3$ content of 46 % in the PHDLC is increased to 49% in the ECHDLC film, as estimated from the above mentioned formula[17]. Electrochemical conversion of $sp^2$C=C into $sp^3$C-H carbons is rarely found in published report [19]. In this report [19], $H^+$ ions in the acidic electrolyte are attracted towards appropriately biased negative $sp^2$C=C carbon cathode for reaction to form $sp^3$C-H carbon. In the present electrochemical conversion of $sp^2$C=C into $sp^3$C-H carbons, the effect of $\mu_e$ being above $E_F$ (see fig.2b) results in negatively charged PHDLC surface and makes the conversion, as described above, more elegant and novel. Raman spectrum gives measurement of bulk material, in general. But in this particular case, since ECHDLC is produced from the surface modification of PHDLC via electrochemical process, the enhancement of PL background (see fig.3a&b) and broadening of PL spectra (see fig.3c) are indicating surface enhancement of Raman signal due to increment of $sp^3$ content onto the surface by about 3%.

4. Concluding Remarks

In the present work, with the objective of hydrogenation of $sp^2$C=C into $sp^3$C-H via covalent bonding during electrochemical process in the PHDLC surface, it is observed that PHDLC surface is capable of undergoing nearly reversible surface redox couple reaction; hydrogenation of $sp^2$C=C into $sp^3$C-H via covalent bonding in 1M KCl (pH=1) electrolyte has been possible due to negative electron affinity of adsorbate ($\chi_{ad}$), causing electron transfer from liquid to PHDLC surface. Hydrogenation has been distinguished from the observed enhancement of Raman and PL spectra in the ECHDLC sample w.r.t that of PHDLC sample. Raman cross-section for the detection of single molecule is observed to be enhanced by several orders of magnitude, when the molecule is adsorbed onto a metal surface. This phenomenon was first discovered by M. Fleishmann in 1974 [20] and thereafter named as surface enhanced Raman signal (SERS) effect [21].Therefore our results on the Raman/PL spectra, seem to be a typical SERS effect. This SERS effect can be used in future, as a probe to distinguish surface modification and hence surface conductance property of PHDLC via electrochemical process.

Acknowledgements

One of the authors (NRR) gratefully acknowledge for funding XI-plan (2007-2012) to create the new experimental facilities at SINP, used in the present work.

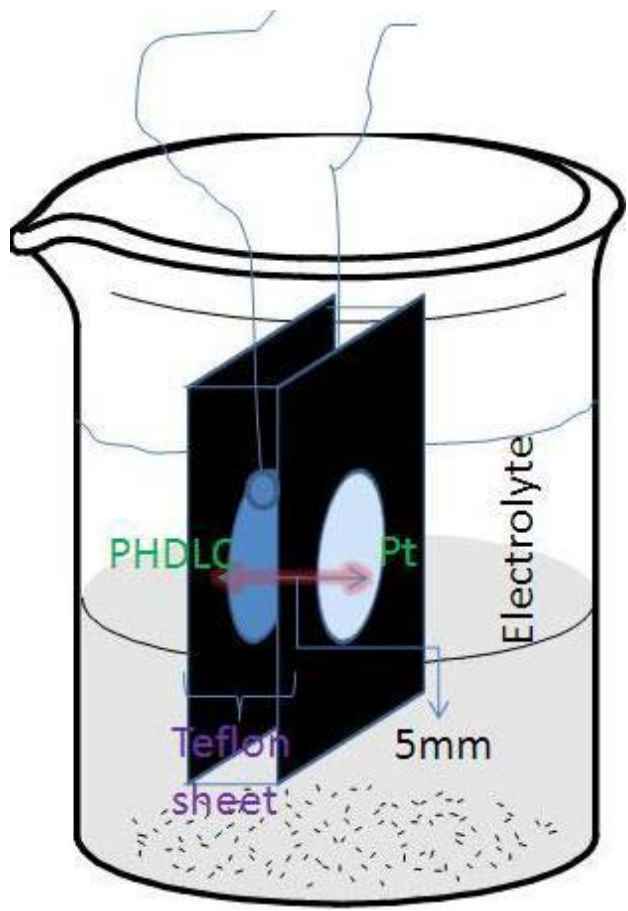

Figure1.  A schematic diagram of the electrochemical cell

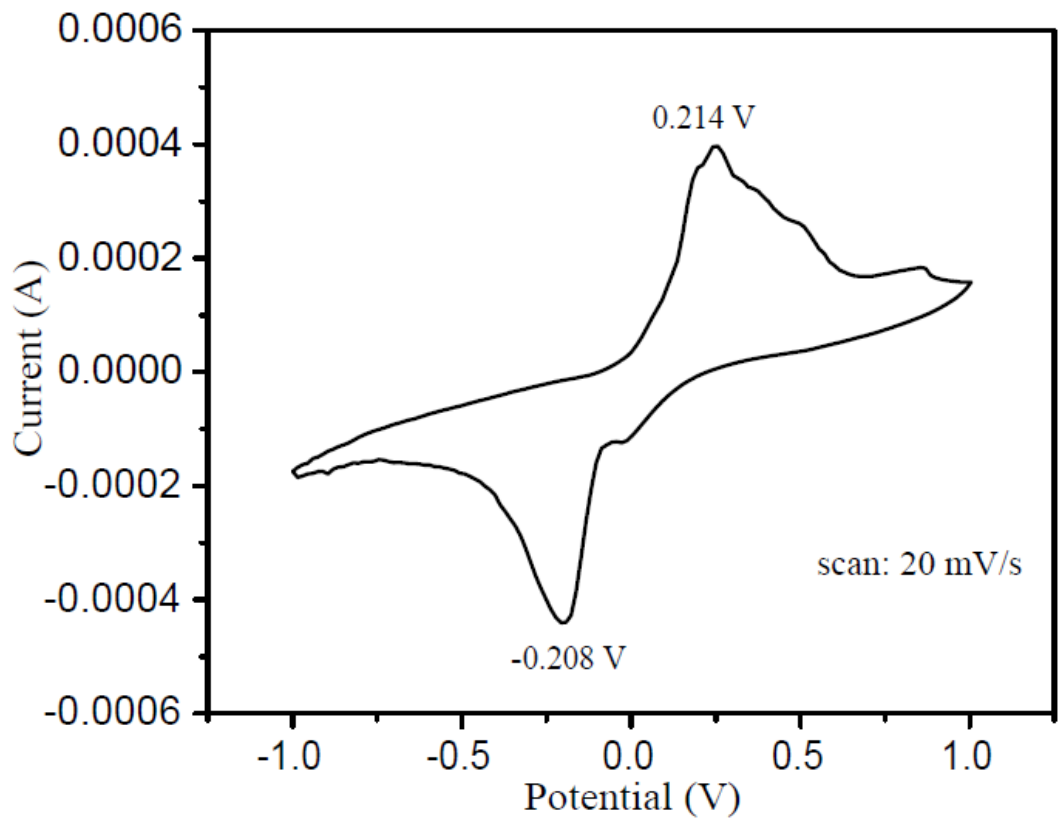

Figure2a  A typical CV measurement of PHDLC film in 1M KCl (pH=7)

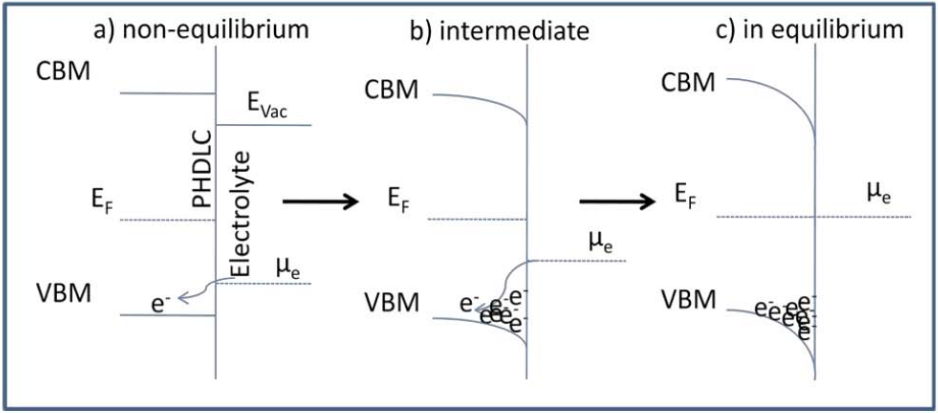

Figure2b  A schematic picture of the electron transfer process between PHDLC surface and electrolyte

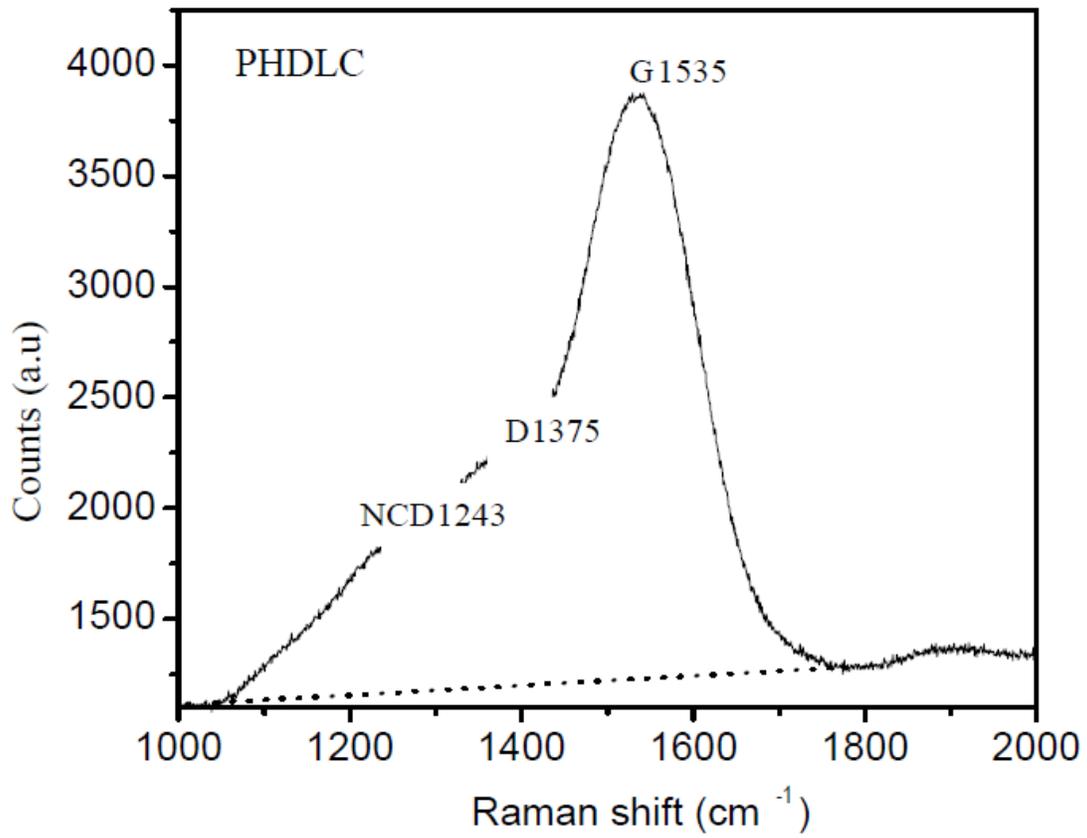

Figure3a The typical Raman spectrum of PHDLC

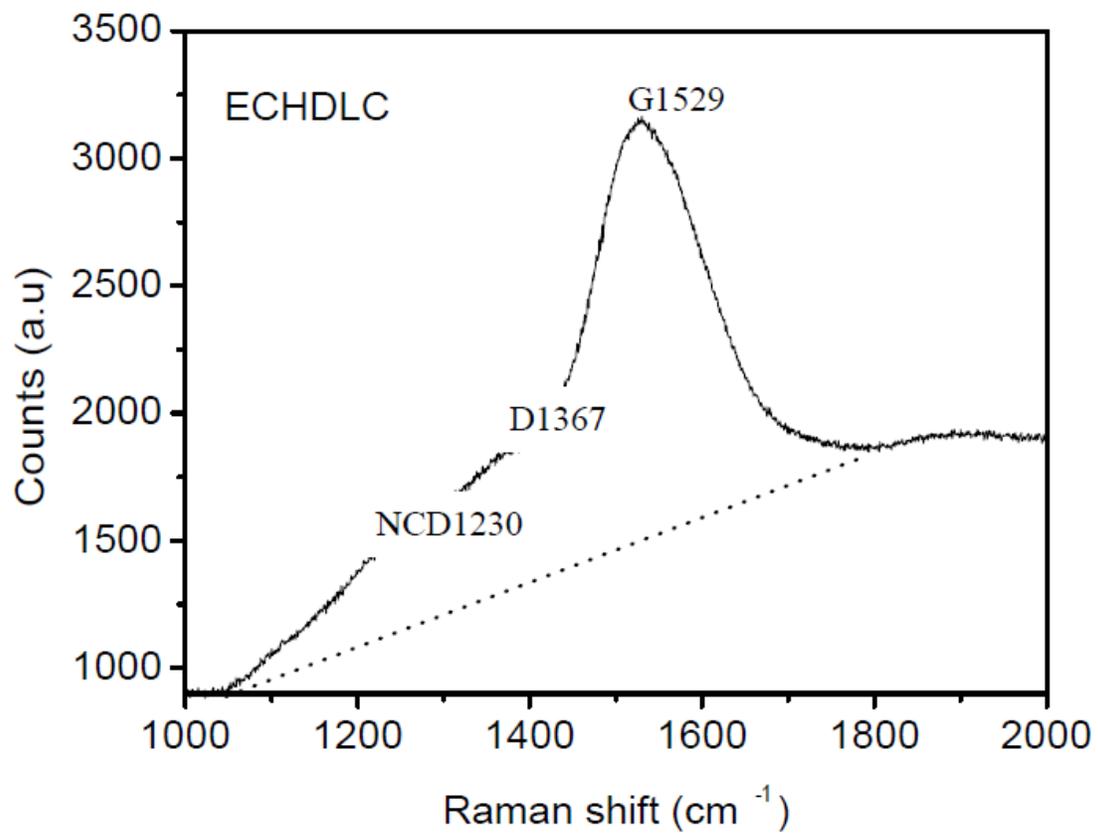

Figure3b The typical Raman spectrum ECHDLC

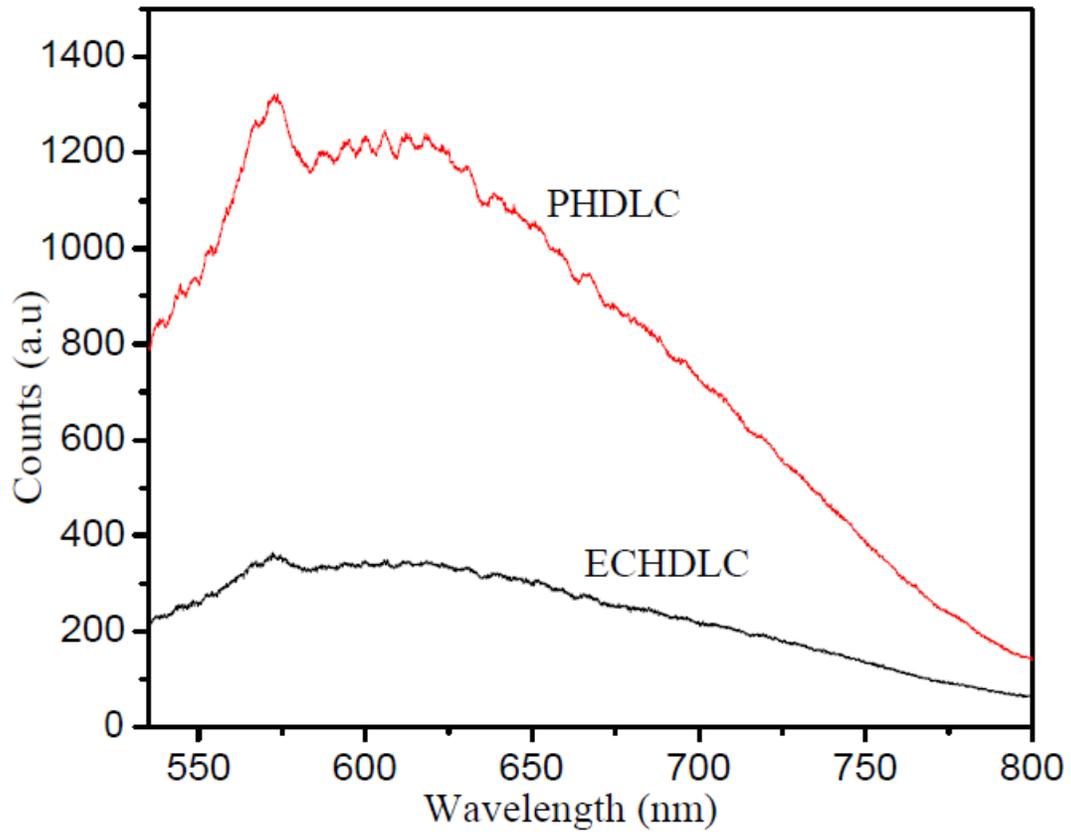

Figure3c PL spectra of PHDLC and ECHDLC